\documentclass[12pt]{article}
\usepackage{epsfig}
\usepackage{amsmath,bm}

\def\beq{\begin{equation}}
\def\eeq#1{\label{#1}\end{equation}}
\def\eeqn{\end{equation}}

\def\beqa{\begin{eqnarray}}
\def\eeqa#1{\label{#1}\end{eqnarray}}
\def\eeqan{\end{eqnarray}}

\let\bar=\overbar

\def\Dslash{\not{\hbox{\kern-4pt $D$}}}
\def\dslash{\not{\hbox{\kern-2pt $\del$}}}

\def\msb{{\bar{\ssstyle M \kern -1pt S}}}

\def\Title#1{\begin{center} {\Large {\bf #1} } \end{center}}
\def\lesssim{\ \hbox{\raise 2pt \hbox{$<$} \kern -13pt
                     \lower 3pt \hbox{$\sim$}}\ }
\def\greatersim{\ \hbox{\raise 2pt \hbox{$>$} \kern -13pt
                     \lower 3pt \hbox{$\sim$}}\ }

\begin{document}
\renewcommand{\thefootnote}{\fnsymbol{footnote}}
\hspace*{11.7 cm} {\small OUTP-07-18-P}

\vspace*{1.2 cm} 

\Title{Hadron structure for x $\ll$ 1 and upcoming 
collider measurements\footnote{Presented 
at the IX Workshop on 
Nonperturbative QCD, IAP, Paris, 4-8 June 2007.}}

\bigskip\bigskip

\begin{center}  

{\it F.~Hautmann\\ 
Theoretical Physics Department, 
University of Oxford,  
Oxford OX1 3NP}
\bigskip\bigskip
\end{center}

\begin{center}
Abstract
\end{center}
\noindent  We discuss theoretical aspects of  
parton distribution functions for very high energy scattering    
in  relation with  
upcoming  measurements in DIS and hadron-hadron collisions.

\section{Introduction}

 The Large Hadron Collider will operate   
with very high parton luminosities. 
As these   rise steeply  for 
decreasing momentum fraction x, 
 a large number of events  sample 
  gluon and sea-quark distributions at  x $\ll$ 1. 
Understanding  
the  theoretical  accuracy in  the determination 
of these distributions is relevant  both to obtain  
reliable predictions for  cross sections of hard processes 
and to investigate   
new aspects of  QCD  physics at very small x.

Current determinations of  parton distribution functions  for 
x $\lesssim 10^{-2} $  largely depend  on deep inelastic
scattering data. We start the
discussion in Sec.~2 by commenting  
on such determinations, with a view in particular     
to the forthcoming longitudinal  structure 
function measurements. 

Most of the  phenomenological pdf analyses 
place cuts on the low-$Q^2$ region, in  order  to 
ensure the applicability of perturbation theory. 
However, a lot of the experimental information on 
x $\ll 1$ physics lies at present with data
that involve  scales of   few GeV.  
In  Sec.~3 we turn to discuss how 
s-channel methods,  
designed to describe high-energy scattering 
 down to lower and lower $Q^2$, 
might be used in the context of the parton picture. 
Measurements of jet production in the DIS current 
fragmentation and central 
region could be exploited in this framework to 
extract  information on the nonperturbative 
parameters of multiple parton scattering.

  We  illustrate in Sec.~4 the use of 
 the above methods  for estimating power 
corrections  
 to structure function's evolution 
 from multiple scattering. We emphasize  the  
distinctive behavior of the results 
in the region of intermediate scales 
just above 1~GeV. 
Through  momentum sum rules and evolution,  
the low-x and low-$Q^2$ region may affect predictions  
 for processes at much higher x and higher  
mass scales. Some additional 
comments on related issues are given 
in Sec.~5. We conclude  
in  Sec.~6.

\section{Flavor-singlet  parton evolution}

Present determinations of parton distribution functions 
are based on  fits to available collider 
data~\cite{alekhin05,cteq06,durham07,durham04}, using  
renormalization-group evolution equations 
to connect measurements 
at different mass scales $\mu$. 
The region x $\ll$ 1 is dominated by 
  flavor-singlet evolution,  
\begin{equation}
\label{flasing}
{d \over {d \ln \mu^2}} \,
\left(\begin{array}{c}
{{ f_S}}\\
{{ f_g}}\\
\end{array}\right)\;
=
\left(\begin{array}{cc}
{
P_{q q }
} & { P_{q g }}\\
{ P_{g q }} & { P_{g g }}\\
\end{array}\right) \;\, \otimes \;\,
\left(\begin{array}{c}
{ f_S}\\
{ f_g}\\
\end{array}\right)\;\;\;\; , 
\end{equation}
where $f_g$ and $f_S$ are the gluon and sea-quark 
distributions, and $P_{i j}$ is the perturbatively calculable 
evolution kernel. 
 The $f_g$ distribution 
in the range x $  \lesssim 10^{-2} $, in particular, is 
relevant  for 
measurements  of production processes dominated by 
gluon fusion at the LHC, including 
heavy flavor production~\cite{baines,heralhc,nason00}   
and Higgs boson production~\cite{heralhc,tev4lhc}. 

At present the extraction  of $f_g$  
  at x $  \lesssim 10^{-2} $   
depends  on DIS data for the 
 $Q^2$ derivative of the  structure function $F_2$ 
  (${\dot F}_2 = d F_2 / d \ln Q^2$),     
\begin{equation}
\label{f2dot}
  F_2 \sim f_S \;\; , \;\;\;\; 
{\dot F}_2 \sim  P_{q g}
\otimes f_g  \ [ 1 + {\cal O} ( \Lambda^2 / Q^2) ] 
 + \ {\rm{quark}} \; \ {\rm{term}} \; . 
\end{equation}
The  kernel $P_{i j}$ used for 
standard  pdf determinations  
is evaluated 
in  fixed order of perturbation theory (NLO and, in 
some of the 
analyses now becoming 
available~\cite{alekhin05,durham07},  NNLO).  As 
indicated by Eq.~(\ref{f2dot}), 
 however, 
the extraction of $f_g$   
is especially  sensitive to 
higher-loop corrections to the gluon $\to$ quark splitting 
kernel $P_{q g}$.  
 In particular, for x $\ll$ 1
logarithmically enhanced contributions  
 $\alpha_s^{ k + 1} x^{-1}  \ln^{ k - 1} x$~\cite{qgx} 
to $P_{q g}$ 
are  
present     for  any $k \geq 1$, and need to be resummed.  
The numerical impact of  terms of this type 
on the global fits is examined e.g. in~\cite{durham04}. 

While the summation of  $\ln x$ terms 
 in the pure-gluon sector 
has been studied extensively 
for quite some time (see  reviews 
in~\cite{heralhc,jeppe06}, and references therein),   
only recently have  the first 
analyses  appeared~\cite{ccss07,thorne,abf07} 
that  implement  both   
 gluon~\cite{flcc}  and quark~\cite{CH94} 
  next-to-leading $\ln x$  corrections to 
  the matrix kernel $P_{i j}$. 
This is theoretically appealing, as quark 
 corrections are required in order to merge consistently 
the x $\to 0$ expansion with the short-distance 
behavior and the renormalization group,  
and opens the possibility for 
 resummed  analyses  to become   directly relevant to 
phenomenology. 

The production of   $b$-quarks  
 at the LHC will be  sensitive to $f_g$ at 
x $  \lesssim 10^{-2} $. Recall   that the 
theoretical uncertainties on 
 NLO predictions for 
the $b$ cross sections, while on the order of a 
few ten percent    at the Tevatron~\cite{bquark}, 
increase to well over a factor of 2 
 at  LHC energies~\cite{baines,nason00,bquark}. 
Improving upon present predictions will likely involve  
a variety of physical 
 effects, from  higher-loop corrections beyond NLO to 
 nonperturbative  processes both in the initial and the 
 final states.    
An improved  treatment of 
x $\to 0$  contributions  for both the pdfs and the 
short-distance cross sections should   help 
understand the sources of the 
 large uncertainties  at the LHC,  
and possibly reduce them. 

The production of Higgs bosons   
may also receive   
sizeable contributions from low-x gluon 
events at the LHC, depending on the Higgs mass range 
 and the kinematical region in the Higgs rapidity and $p_T$. 
 See e.g.~\cite{hgsx}  
for studies of high-energy  effects  
on the accuracy of 
 Higgs boson cross sections.

The x $\ll$ 1 region will be probed experimentally by 
  the forthcoming 
measurements of the longitudinal  
structure function $F_L$~\cite{hera-fl}. 
A recent  phenomenological study  of $F_L$ may be 
found  in 
Ref.~\cite{thorne}. This presents a thorough    
comparison of fixed-order   predictions 
 through NNLO~\cite{moch} and  
 predictions including the  $\ln x$ resummation for 
 the $F_L$ coefficient functions~\cite{CH94}.   
The analysis~\cite{thorne} shows that i) the 
NLO and NNLO corrections to $F_L$ are 
  large, as observed in~\cite{moch},
and lead to strong instabilities  
 causing perturbation theory to be badly 
behaved at low x; ii)    
 resummed results improve this behavior,  and allow 
one to obtain  more reliable  
 theoretical predictions throughout a wider range in the 
kinematical variables. 
This is potentially significant, since the $F_L$ measurements 
will provide an independent observable, to  be combined with 
${\dot F}_2$,  to perform 
 a complete  flavor-singlet analysis and 
 probe  the accuracy 
of the theory at  x~$ < 10^{-2}$ 
 more stringently  than ever so far.

A different, but possibly related  issue concerns 
  the initial conditions 
 at low mass scales for pdf evolution.  
Working to  
NNLO,  Ref.~\cite{dortm} observes that 
    rather different  features 
 than in~\cite{thorne,moch,mst06} 
are obtained 
in  the fixed-order predictions for $F_L$ 
 if   different assumptions are made~\cite{dortm, dortm98} for the 
initial conditions.  
We note that this marked dependence  on the initial 
condition may  be  related to using a fixed-order 
truncation of perturbation theory that is  
 not well-behaved at low x, and could likely   
 be reduced in the improved  theory including  resummation.

On the other hand, 
 while resummed evolution schemes 
 provide  better theoretical control on the 
x $\ll$ 1 region~\cite{ccss07,thorne,abf07}, 
note that the coefficients themselves of 
resummed perturbation theory  signal 
 the onset of  dynamics  beyond the $\alpha_s$ expansion. 
One way to see this is to examine the 
singularity structure  
of resummed coefficients  in the plane of the 
effective x $ \to 0$ anomalous dimension 
$\gamma_\omega (\alpha_s)$~\cite{rsings}. 
The strong  branch-point  
singularities  in gluon-channel coefficients~\cite{rsings}  
 $R_\omega (\alpha_s)  \sim 
1 / \sqrt{ 1/2 - \gamma_\omega (\alpha_s)}$, with 
$\gamma_\omega \to 1/2$ for x~$ \to 0$, imply 
that  effects  beyond the  perturbation expansion are to be 
included in order to fully describe the 
high-energy limit (see also discussion in~\cite{cc97}).

In Eq.~(\ref{f2dot}) we have indicated explicitly 
contributions to 
${\dot F}_2$ suppressed by powers of $1 / Q^2$. 
Because   present  DIS  data that are   relevant to extract 
the gluon density for x $ < 10^{-2}$ do not have very high 
$Q^2$, these power corrections 
may be expected to have non-negligible effects on the 
 estimate of the 
theoretical accuracy on  pdf's for the LHC. 
In the next 
sections we turn to methods for  
the subset of power-like contributions 
 that comes    from graphs with 
multiple gluon  scatterings.

\section{Relating parton approach and 
s-channel approach}

Methods to take account of multiple scatterings are 
based on the s-channel picture of  deep 
inelastic collisions. See~\cite{bj90s} for an introduction 
to the physical picture, 
and~\cite{ianmue07,kovwei,gelvenu07,golrev}  
for motivation from high-density QCD. 
The main advantage of this approach 
is that it gives a formulation 
that can be used down to small $Q^2$, 
 incorporating nonperturbative 
physics at very high energies 
in Wilson-line operator matrix elements.  

The s-channel approach leads to a different picture of 
the hard collision than the parton approach, as it 
works in 
 a different reference frame, and uses different 
 degrees of freedom. 
Nevertheless, the two pictures are not incompatible. 
In the region where their domains of validity overlap, they 
must describe the same physics. 
Ref.~\cite{hs07} presents a framework to 
connect the two pictures. 

\begin{figure}[htb]
\begin{center}
\epsfig{file=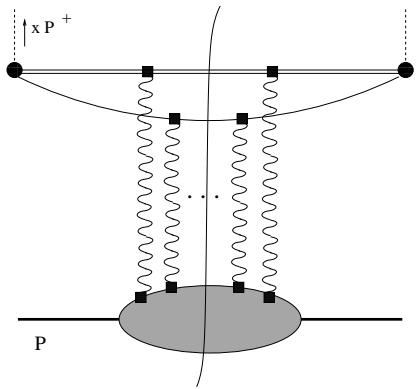,height=2.0in}
\caption{Quark distribution function 
in the s-channel picture.}
\label{fig:quapdf}
\end{center}
\end{figure}

The method~\cite{hs07} is  based on
constructing explicitly  an s-channel
representation for the renormalized parton 
distribution function
   in terms of
 Wilson-line matrix elements.  In    this
  representation  the quark distribution $f_q$
is given by  the coordinate-space 
convolution (Fig.~\ref{fig:quapdf})  
\begin{equation}
\label{convfq}
x f_q ( x , \mu) = \int {d{\bm z}} \int {d{\bm b}} \
u (\mu ,  {\bm z} ) \ \Xi( {\bm z}, {\bm b})  - UV   
\;\; ,
\end{equation}
where 
$\Xi$ is the hadronic  matrix element 
of  eikonal-line
operators,
\begin{equation}
\label{xidef}
\begin{split}
\Xi( {\bm z}, {\bm b}) =
\int [ d P^\prime ] \
\langle P'|\frac{1}{N_c} \ {}& {\rm Tr}\{1 -
V^{\dagger}( {\bm b} + {\bm z}/2)\,
V({\bm b}-{\bm z}/2)
\} |P \rangle \hspace*{0.2 cm} ,
\\&  V({\bm z}) = {\cal P}\exp\left\{
-ig\int_{-\infty}^{+\infty}dz^- { A}^+_a(0,z^-,{\bm z}) t_a
\right\}   \;\;  ,
\end{split}
\end{equation}
${\bm z}$ is the transverse separation between the eikonal lines,
${\bm b}$ is the impact parameter,
and the
function $u (\mu ,  {\bm z} )$ is evaluated explicitly
 in~\cite{hs07} at one loop
using the $\overline {\rm MS}$   scheme for
the renormalization of the
ultraviolet divergences ${\bm z} \to 0$.

The  convolution 
 (\ref{convfq}) expresses the fact that for x $\ll$ 1 
the operator defining the quark 
distribution function creates the partonic system, made of a  
color-triplet eikonal line and an antiquark, 
at large  longitudinal distances 
$y^- \approx 1 / (x P^+)$ far outside the target~\cite{bj90s}. 
The $\overline {\rm MS}$ 
result~\cite{hs07} can  be recast in a  
physically more transparent form 
in terms of a cut-off on the ${\bm z}$ 
integration region (see also~\cite{hs-pho07}), as long as 
the scale $\mu$ is sufficiently large compared to the   
inverse hadron radius: 
\begin{equation}
\label{eq:renormalizedfq}
 xf_{q/p}(x,\mu) 
=   
\frac{N_c}{3 \pi^4}\
\!\int\! {d{\bm b}} \ { {d{\bm z}} \over  { {\bm z}^4 }  }  \  
\theta({\bm z}^2\mu^2 > a^2) \
\Xi({\bm z} , {\bm b} )  \;\;\; , 
\end{equation}
where $a$ is a renormalization scheme dependent 
coefficient~\cite{hs07}, 
\begin{equation}
\label{eq:aresult}
a = 2 e^{1/6 - \gamma} \approx 1.32657 \;\;  , 
\end{equation}
with $\gamma$ the Euler constant.   
 The Wilson-line  matrix element 
 $\Xi({\bm z}, {\bm b})$ 
receives contribution from both long  distances and 
short distances. 
At small ${\bm z}$ it 
  can be treated by a short distance 
expansion. 
By using renormalization-group 
evolution equations, the 
 leading term   in the expansion  can be  
  related to    a 
well-defined integral of the 
 gluon distribution function~\cite{hs07}. 
 At large distances ${ \bm z}$, $\Xi$ 
  has to be  fit to 
data, or parameterized by models 
consistently with bounds from unitarity and 
saturation~\cite{golrev,mue99}.  
 
Besides parton distribution functions,  
  the 
s-channel representation discussed above 
 is potentially interesting for jet production. 
To this end, we note that 
Eqs.~(\ref{convfq}),(\ref{eq:renormalizedfq})
 imply introducing  a cut-off 
in rapidity~\cite{hs07} in order 
 to separate slow and fast partons in 
the picture of Fig.~\ref{fig:quapdf}, and to  
factor out  the Wilson-line matrix element. 
The dependence on the method of carrying out this 
separation   enters in higher orders for the quark 
distribution, and in leading order for the gluon. 
Both the quark and the gluon channel contribute to 
the jet structure of the final states. 
Nevertheless, 
if one considers  jet 
cross sections  
 by fixing e.g. the total minus momentum of the dijet system, 
the sensitivity to the rapidity cut-off 
  is suppressed,  
as long as the jets are produced sufficiently far from the 
fragmentation region of the target. 
 Then the application of the 
s-channel method 
to  measurements of jet production 
 will give 
 rather direct 
 information on the contribution of  multiple 
 parton scatterings. 
 
 Note that current 
shower Monte-Carlo generators 
include model parameterizations of 
multiple interactions in order to produce realistic event 
simulations, see e.g.~\cite{bartalini,sjoskands}.
 In this respect, 
  jet 
  rapidity distributions  
 in the DIS current fragmentation and central regions, once 
 analyzed 
using the s-channel picture,  
 are potentially useful  in order to constrain the 
nonperturbative parameters of 
such models.

The physical picture described in this section 
relies 
 on the large separation in lightcone coordinate 
between  creation of the partonic system 
and interactions with the target. 
This separation is of order a hundred fermi 
in the proton rest frame for   x~$ \lesssim 10^{-2}$ (see 
remark after Eq.~(\ref{xidef})).  
The condition  for 
the applicability of the approach is satisfied 
 in the case of 
 high-energy collisions of large nuclear 
targets as well. The approach should be relevant   
   for the physics of 
high-$p_t$ probes in 
heavy ion collisions at the LHC 
and of 
 nuclear parton distributions~\cite{accar}.

\section{Power corrections from the s-channel}

The  representation  (\ref{convfq}), 
 evaluated in a  well-prescribed 
renormalization scheme,  is the key ingredient that 
 allows one  to 
relate~\cite{hs07,plb06} results of s-channel 
calculations for   structure 
functions~\cite{ianmue07,golrev,mue99}   to the OPE 
factorization, 
\begin{equation}
\label{opefac}
F_2 = C \otimes f + { 1 \over Q^2}  \
C^{(4)} \otimes f^{(4)} + \dots \;\; , 
\end{equation}
and,  in particular, to   identify  
power-suppressed contributions to the $Q^2$ evolution of  
 $F_2$ 
of the form 
\begin{equation}
\label{df2}
 {\dot F}_2 \simeq P_{q g}
\otimes f_g \ \left[ 1 + \delta
\right] + \ {\rm{quark}} \; \ {\rm{term}} \;
\hspace*{0.2 cm}  ,
\hspace*{0.2 cm}
\delta \simeq \sum_{k \geq 1} \ a_k \ (\alpha_s  
\ {1 \over x^\beta} \
{\Lambda^2 \over Q^2})^k \;\; .  
\end{equation}
The enhanced x $ \to 0$ behavior in the power 
correction $\delta$ in Eq.~(\ref{df2}) 
is produced from graphs with multiple 
gluon scatterings, and 
is consistent 
with observations of approximate geometric scaling 
in low-x data~\cite{geoscal}.

Refs.~\cite{plb06,morio07} introduce moments $\lambda^2$ 
 of the matrix element $\Xi$   in Eq.~(\ref{xidef}), 
\begin{equation}
\label{moments}
\lambda^2 (- v)   = {1 \over \Gamma ( v ) }
\int   {{d{\bm z}} \over {\pi {\bm z}^2}}
\ ({\bm z}^2 )^{v - 1} \
 \int d {\bm b} \ \Xi( {\bm z}, {\bm b}) \; \; ,  
\end{equation}
and express the 
$1 / (Q^2 )^k$ correction in terms of $\lambda^2 (k)$ times 
 coefficients computable as functions of 
$\alpha_s$, $x$ and $\ln Q^2$, schematically in the form 
\begin{equation}
\label{schemat}
{{d F_2} \over { d \ln Q^2 }} =
\left( {{d F_2} \over { d \ln Q^2 }} \right)_{\rm{LP}}
+ \sum_{k=1}^\infty \ R_k \ {\lambda^2 (k) \over {( Q^2 )^k}} \; \; . 
\end{equation}
The first term in the right hand side is the leading-power 
parton result, and the 
 moments $\lambda$ 
 in the subleading terms 
 are dimensionful nonperturbative 
parameters, to be determined from comparison with 
experimental data. 
Observe that any number of rescatterings contribute to the 
moment $\lambda^2 (k)$ through the eikonal 
operators (\ref{xidef}). 
Thus the correction of order $1 / (Q^2 )^k$ 
receives contribution from the exchange of 
arbitrarily many 
gluons. This is characteristic of the small-x power correction 
 and 
can be contrasted with counting rules arguments, valid at 
large x, that link the power in $Q^2$ with the number of 
parton lines exchanged in the t-channel.

\begin{figure}[htb]
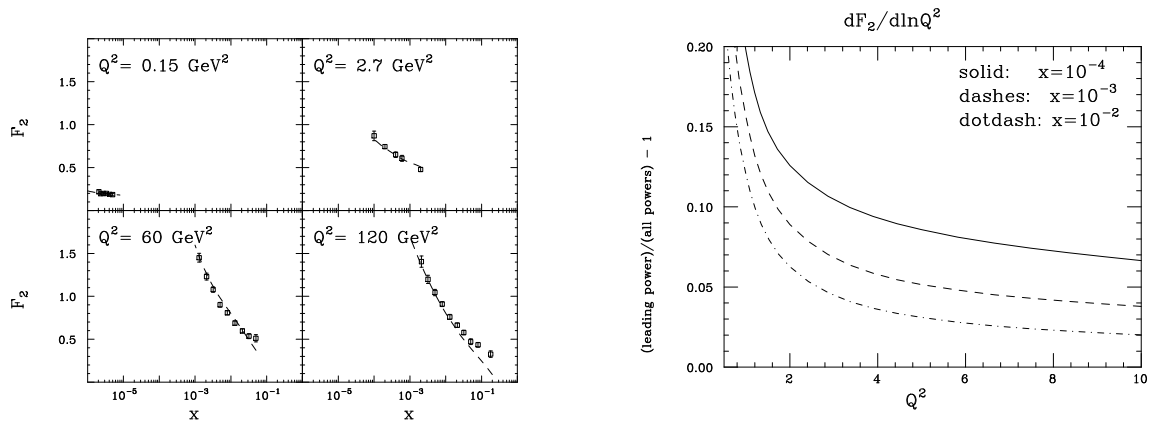

\vspace{55mm}
\includegraphics[scale=0.3,angle=90,bbllx=30,
bblly=400,bburx=-440,bbury=760]{dat_fourpan.ps}
\includegraphics[scale=0.32,angle=90,bbllx=70,
bblly=760,bburx=-400,bbury=1120]{allpow_lin.ps}
\caption{(left)  The result of fitting  
 the  $\lambda^2$  parameters 
to the  data\protect\cite{zdata}; (right)
power corrections to
$d F_2 / d \ln Q^2$ versus $Q^2$ at different values 
of x\protect\cite{plb06}. 
}
\label{fig:withtwo}
\end{figure}

  The result of 
 determining the nonperturbative 
 $\lambda$   parameters  
  from    $F_2$ data~\cite{zdata} at  both low and high 
$Q^2$ is shown in  
Fig.~\ref{fig:withtwo} in the  left hand side plot~\cite{plb06}.   
 The corresponding 
 power correction    
is  plotted on the right hand side of   
Fig.~\ref{fig:withtwo}. Here 
the correction   is 
normalized to the full answer and multiplied by $-1$. 

The analysis~\cite{plb06} indicates that 
with physically natural choices of the parameters in the 
nonperturbative matrix elements 
in Eqs.~(\ref{xidef}),(\ref{moments})  one can achieve 
a sensible description of data for x 
$ < 10^{-2}$  in a wide  range of $Q^2$  
 and  still have moderate power corrections 
to $d F_2 / d \ln Q^2$. Corrections 
turn out to be negative and 
below  20 $\%$ for 
x~$ \greatersim 10^{-4}$ and  $Q^2 \greatersim 1$ GeV$^2$. 
This observation suggests that the power expansion is 
not breaking down, and should still work at least to 
the values of x considered in Fig.~\ref{fig:withtwo}. 

However, we also see from Fig.~\ref{fig:withtwo} 
 that for small x the corrections 
fall off slowly in the 
region of medium $Q^2$, $ Q^2 \simeq 1 - 10$ GeV$^2$, 
behaving effectively like  $1/Q^\nu$ 
with
$   \nu $ close to 1~\cite{plb06}.  For 
instance, one has $ \nu \simeq 1.2$ for the curve x 
$ = 10^{-3}$ in the right-hand side plot of Fig.~\ref{fig:withtwo}.  
As a consequence of the slowly decreasing behavior, 
the power corrections  stay on the order of 
$ 10 \%$ up  to $Q^2$ of a few GeV$^2$ for x $ 
 \lesssim 10^{-3}$. 
This  slow fall-off    differs 
from parameterizations of higher twist commonly   used 
in global analyses (see e.g. Ref.~\cite{durham04}), and 
may be relevant 
for phenomenology as it affects the medium $Q^2$ region of the   
data that are useful 
  to extract $f_g$ at low x. 

\section{Further comments}

The  region 
of low x and low-to-medium  $Q^2$ 
  influences   
predictions at much higher x and mass scales through 
momentum sum rules and evolution.  We give below some 
further comments. 
 
It is worth noting that the 
 biggest contribution to the power correction to 
${\dot F}_2$ in~\cite{plb06} comes from the 
longitudinal component. The 
 derivative $d F_T / d \ln Q^2$ has 
smaller power corrections than $d F_2 / d \ln Q^2$. 
This provides additional motivation for the  
separate measurement of $F_L$~\cite{hera-fl}.  
Similar conclusions are reached 
from a different perspective 
based on the  
fits~\cite{durham04,mst06}, investigating 
the effect of power-like terms on global analyses 
at both high x and low x.     
(These  fits also include parameterizations of 
power effects from self-energy graph  
insertions. We recall here that only  
relatively few   results are known as yet 
on such effects   
in flavor-singlet observables.  
See~\cite{gardiweig} for a recent study, and 
early discussions in~\cite{renflav}.)

The curves in Fig.~\ref{fig:withtwo} are obtained using 
NLO parton distributions. It is natural to expect a change 
in the power correction when going from NLO to NNLO. 
We note that the decrease in the 
low-x gluon at NNLO~\cite{durham07} is consistent with the 
possibility that NNLO partons  give 
 smaller power corrections.

Observe that 
in the calculation described in Sec.~4 the slow fall-off with 
$Q^2$ results from summing the terms proportional to the 
moments  $\lambda$  in Eq.~(\ref{moments}).  
These in turn are obtained from expanding the 
s-channel answer in powers of $1 / Q^2$, and enforcing 
consistency with the standard parton framework. 
It will be of interest to compare the $Q^2$ behavior 
 found here 
with the behavior due to  
 the anomalous dimensions 
that result from  nonlinear evolution equations   
  such as those, e.g., used 
  in~\cite{ianmue07,kovwei,gelvenu07,gardiweig}. 

It will also be of interest to investigate the 
relation of the result (\ref{moments}),(\ref{schemat}) 
 for the power  correction  
with the   x-rescaling form   
proposed in~\cite{qiuvitev} 
 and applied to nuclear targets. 
 This should likely involve 
 the finite lightcone-time  cut-off that enters in  
  the high-energy eikonal approximation  
  (see e.g.~\cite{hs00}), reflecting 
the fact that the eikonalized projectile-target 
interactions  do not spread out to arbitrarily large 
times in the far past and the far future.

 The behavior observed in Fig.~\ref{fig:withtwo}   suggests  
that the power corrections  
may be characterized by 
a  nonperturbative scale 
substantially larger than $\Lambda_{\rm{QCD}}$. 
  It is possible that this 
can be related to 
the dynamical cut-off on large transverse 
distances ${\bm z}$ 
 imposed by 
unitarity   requirements 
(``black disc" limit)  on the 
correlator $\Xi$ in Eq.~(\ref{xidef}). To pursue this,  
further studies of the moments (\ref{moments}) 
are warranted. Also, the 
analogue of these 
moments for  eikonal operators (\ref{xidef}) 
in the adjoint representation will be 
relevant for processes 
coupled directly to the gluon distribution. 
An especially interesting case  is that of 
diffractive DIS events~\cite{abra}, 
where  cross sections depend
 quadratically on the $\Xi$ 
correlator, and gluonic contributions typically dominate 
quark contributions by one order of magnitude.   
 See comments in~\cite{morio07},   
 and references therein, on the   possible 
 role of  the ``blackness" cut-off in hard-diffractive data. 

\section{Conclusion}

The partonic structure of 
protons and nuclei will be probed 
for small values of x at the LHC.  
   DIS measurements of  
longitudinal $F_L$, 
  expected in the coming year, 
will provide  very valuable, new  experimental input. 
As discussed in Sec.~2, 
improved theoretical tools for 
the evolution of parton distributions 
are starting   to become available   
that include  the results of 
next-to-leading $\ln x$ 
 resummation 
 for both gluon {\em and}  
quark channels, as is 
required by consistency with the short-distance 
behavior and the renormalization group.

Much of the present  information on the 
region x $ < 10^{-2}$  comes from deep inelastic data 
involving scales of few GeV.  
Physics beyond   the perturbation expansion
is likely to play a role in this regime. 
 The physical picture 
of hard collisions that is designed   to  
go down to lower and lower $Q^2$ when x is small 
is the s-channel picture.   
A method to use results of s-channel calculations in 
the context of the parton framework is discussed in 
Sec.~3, and applied in  Sec.~4 to 
corrections to structure function's 
evolution suppressed by powers of 
$1 / Q^2$ but enhanced as x $\to 0$.   

The results presented build a physical picture 
of the sea-quark distribution for very high energies, 
that allows one to discuss quark saturation. 
Nonperturbative power-like effects are expressed 
in terms of moments of  Wilson-line eikonal correlators.  
The picture suggests the possibility of using measurements 
of  jet leptoproduction in the  intermediate rapidity region 
to measure effects of  parton rescatterings. 

\bigskip
\noindent {\bf Acknowledgments}. I thank the 
workshop's  organizers and staff  for the 
kind invitation. 
I thank D.~Soper for discussions and for careful 
reading of the manuscript.

\end{document}